\documentclass[11pt]{article}
\usepackage{graphicx,cite,lineno,amsmath,amssymb,amsfonts,color}
\usepackage{multirow}
\newcommand\Sl{\tilde l}
\newcommand{\pt}{$p_{T}$}
\newcommand{\sq}{$\sqrt{s}$}

\newcommand{\fbinv}{fb$^{-1}$}

\usepackage[symbol]{footmisc}

\usepackage[a4paper,top=3cm,bottom=3cm,left=2.5cm,right=2.5cm,bindingoffset=0mm]{geometry}

\begin{document}

\titlepage

 \begin{flushright}                                                    
 IPPP/18/103                                                                                                    
  \end{flushright}   

\vspace*{0.5cm}
\begin{center}
  {\Large\bf
LHC Searches for Dark Matter in Compressed Mass Scenarios:\\ \vspace{0.3cm} Challenges in the Forward Proton Mode}
  \\  
  \vspace*{1cm}

{\sc 
L.A. Harland-Lang$^{1}$%
\footnote[1]{email: lucian.harland-lang@physics.ox.ac.uk}%
, V.A.~Khoze$^{2,3}$%
\footnote[2]{email: V.A.Khoze@durham.ac.uk}%
, M.G.~Ryskin$^{3}$%
\footnote[3]{email: ryskin@thd.pnpi.spb.ru}\\%
and M.~Tasevsky$^{4}$%
\footnote[4]{email: Marek.Tasevsky@cern.ch}%
}
\vspace*{0.5cm}

{\small\sl
  $^1$Rudolf Peierls Centre, Beecroft Building, Parks Road, Oxford, OX1 3PU, UK

\vspace*{0.25cm} 

$^2$IPPP, Department of Physics, University of Durham, 
Durham, DH1 3LE, UK

\vspace*{0.25cm} 

$^3$Petersburg Nuclear Physics Institute, NRC ``Kurchatov Institute'', Gatchina,
St.~Petersburg, 188300, Russia
\vspace*{0.25cm}

$^4$Institute of Physics, Czech Academy of Sciences, 
CS-18221 Prague 8, Czech Republic
}

\vspace*{1cm}                                                    
                                                    
\begin{abstract}                                                    
\noindent We analyze in detail the LHC prospects at the center-of-mass enery of \sq\ = 14~TeV for charged electroweakino searches, decaying to leptons, in compressed supersymmetry scenarios, via exclusive photon-initiated pair production. This provides a potentially increased sensitivity in comparison to inclusive channels, where the background is often overwhelming. We pay particular attention to the challenges that such searches would face in the hostile high pile--up environment of the LHC, giving close consideration to the backgrounds that will be present. The signal we focus on is the exclusive production of same-flavour muon and electron pairs, with missing energy in the final state, and with two outgoing intact protons registered by the dedicated forward proton detectors installed in association with ATLAS and CMS. We present results for slepton masses of 120--300 GeV and slepton--neutralino mass splitting of 10--20 GeV, and find that the relevant backgrounds can be controlled to the level of the expected signal yields. The most significant such backgrounds are due to semi--exclusive lepton pair production at lower masses, with a proton produced in the initial proton dissociation system registering in the forward detectors, and from the coincidence of forward protons produced in pile-up events with an inclusive central event that mimics the signal. We also outline a range of potential methods to further suppress these backgrounds as well as to enlarge the signal yields.

\end{abstract}                                                        
\vspace*{0.5cm}                                                    
                                                    
\end{center}                                                    

\section{Introduction}
One of the main goals of the physics program at the LHC and future colliders is
the search for  beyond the Standard Model (BSM) physics. One well--motivated, and much
explored, candidate among the existing BSM scenarios is the Minimal Supersymmetric
Standard Model (MSSM), see e.g.~\cite{Haber:1984rc,Martin:1997ns,Djouadi:1998di,Baer:2006rs} for reviews. This in particular  offers a natural candidate for cold Dark Matter (DM), the Lightest
Supersymmetric Particle (LSP), which is the stable lightest neutralino, 
$\widetilde{\chi}^0_{1}$, see e.g. \cite{Goldberg:1983nd,Ellis:1983ew}. 

A possibility that has received significant recent attention in the context of LHC and future collider searches is the electroweak pair production of $R$-parity conserving
states in compressed mass scenarios of supersymmetry (SUSY). That is, where the mass difference between the
heavier state (e.g. the chargino, $\widetilde{\chi}^\pm $, or slepton, $\tilde l(g)$) and the LSP
$\widetilde{\chi}^0_{1}$ is small, see for instance \cite{LeCompte:2011cn,Baer:2014yta,Moortgat-Picka:2015yla,Khachatryan:2016mbu,CMS:2016zvj,Golling:2016gvc,Lehtinen:2017vdt,Khoze:2017ixx,Aaboud:2017leg,Sirunyan:2018lul,Sirunyan:2017lae,Sirunyan:2018iwl,Bagnaschi:2017tru}. Such models are well motivated by cosmological
observations and naturalness considerations~\cite{Barbieri:1987fn,deCarlos:1993rbr,Giudice:2010wb,Han:2013usa,Schwaller:2013baa} as well as $(g-2)_{\mu}$ phenomenology~\cite{Ajaib:2015yma,Ghosh:2016uwl,Kobakhidze:2016mdx,Endo:2017zrj}. 
In particular, the small mass splitting between the DM state and its
charged partner plays a crucial role in bringing the dark matter abundance down
into agreement with the observed value through co-annihilation mechanism~\cite{Griest:1990kh,Baker:2015qna,Edsjo:1997bg}.

As charginos and sleptons only interact electroweakly,
their direct production cross sections at the LHC are quite small and
correspondingly the LHC discovery potential and current experimental
bounds are substantially weaker in comparison to other SUSY states~\cite{ATLASsearch,CMSsearch}. Moreover, searches in the compressed mass region 
via standard inclusive channels are experimentally very challenging, in particular because the SM $WW$ background
 produces a very similar final state, and therefore contaminates any potential signal. A further difficulty is
that the visible decay products have low momenta and therefore often do not pass detector
acceptance thresholds. In order to trigger on such events, generally the presence of an additional jet or photon due to initial state radiation
is required, providing the final-state particles with a boost in the
transverse plane, and thus generating a  large missing transverse
momentum.

With these challenges in mind, the potential to search for these comparatively light charged SUSY particles via photon--initiated production in hadron collisions has been widely discussed over the past few decades~\cite{Ohnemus:1993qw,Piotrzkowski:2000rx,Schul:2008sr,Khoze:2010ba,HarlandLang:2011ih}. One clear benefit of considering this channel is its model independence, in sharp
contrast to many other reactions. That is, the production cross sections are directly predicted in terms of the electric charges of the relevant states.
Now, the development of the forward proton
detectors (FPD) at the LHC allows us to perform a wide program of such searches~\cite{Albrow:2008pn,N.Cartiglia:2015gve,Kepka:2008yx,Royon:2016hdx,Harland-Lang:2016kog,Khoze:2017igg,Baldenegro:2018hng,Harland-Lang:2015cta,Harland-Lang:2018iur,SuperCHIC}. In particular, dedicated AFP~\cite{AFP,Tasevsky:2015xya} and CT--PPS~\cite{CT-PPS} FPDs have recently been installed in association with both ATLAS and CMS, respectively. The purpose of these near-beam detectors is to measure intact protons arising
at small angles, giving access to a wide range of Central Exclusive Production
(CEP) processes
\begin{equation}
pp\to p~+~X~+~p\ ,
\end{equation}
where the plus sign indicates the presence of the Large Rapidity Gaps (LRG) between the produced state and outgoing protons.
The experimental signature for the CEP of electroweakinos is then the 
presence of two very forward protons that are detected in the FPD and two leptons
from the
$\tilde l(g)\to l+\widetilde{\chi}^0$ decay 
 whose  
production vertex is indistinguishable from the primary vertex measured in the
central detector. The well--defined initial state and presence of the tagged outgoing protons provides a unique handle, completely absent in the inclusive case, that is able to greatly increase the discovery potential.

The FPDs consist of tracking and timing detectors
which are inserted in Roman Pots. They are placed at roughly 220\,m from the
interaction point (IP) on both sides of the ATLAS and CMS detectors. Their
acceptance in the fractional momentum loss of the intact protons, $\xi$, is
approximately $0.02 < \xi < 0.15$ at the nominal accelerator and beam
conditions, which corresponds to the unprecedented acceptance of 250 to 1900~GeV
in the invariant mass of the central system $X$ when both protons remain intact and are registered in the detectors.
An important advantage of these detectors is their excellent $\xi$ resolution, which permits a very precise measurement of the
mass of central system, at the percent level.
In the case of slepton pair production  this allows a precise measurement of
the slepton pair and
then, keeping the average di--lepton mass low, this can provide a reasonably accurate determination of the mass of the LSP DM candidate.

In this paper we study in detail the LHC prospects for searching for such exclusive slepton pair production in compressed mass scenarios at the center-of-mass energy of \sq\ = 14~TeV.
Such a possibility was first discussed in~\cite{Khoze:2017igg}  (see also~\cite{vak,vak1,vak2,Marektalk}), and has more recently been considered in~\cite{Beresford:2018pbt}.
Here, we perform for the first time a systematic analysis of the various
challenges and sources of backgrounds that such studies must deal with, a serious consideration of which is essential to assess the potential of these exclusive channels.
In particular, as well as the irreducible exclusive $WW$ background, we also consider the reducible backgrounds from semi--exclusive lepton pair production, where a proton produced in the initial proton dissociation registers in the FPDs, and the pile--up background where two soft inelastic events coincide with an inelastic lepton pair production event. As we will see these two formally reducible backgrounds are expected to play a significant role during nominal LHC running conditions, and require close examination.

The outline of this paper is as follows. In section~\ref{sig} we consider the signal and present the event selection procedure. In section~\ref{QEDbg} we discuss the photon--initiated backgrounds, due both to exclusive $WW$ production and semi--exclusive low mass lepton pair production. In section~\ref{CEPbg} we discuss the semi--exclusive QCD--initiated background due to low mass jet and meson pair production. In section~\ref{ND} we discuss the background from inelastic events in combination with independent pile--up interactions. In section~\ref{FY} we summarize our numerical results for the expected event yields. Finally, in section~\ref{sum} we conclude, while in appendix~\ref{flux} we discuss the
treatment of inelastic photon interactions and in appendix~\ref{treg} we describe how the probability of a proton hit in the FPD due to a proton dissociation system is calculated.

\section{Signal cross section and selection cuts}\label{sig}

As in~\cite{Aaboud:2017leg} we confine ourselves to a simplified SUSY
model and consider the direct pair production of smuons and selectrons
$\Sl_{L,R}$ ($l=e,\mu$) only, where the subscripts $L, R$ denote the left- and
right-handed partner of the electron or muon. The four sleptons are assumed to be
mass degenerate and to decay with a 100\% branching ratio into the
corresponding SM partner lepton and $\widetilde{\chi}^0_1$ neutralino. We consider slepton masses in the 120--300 GeV region, in order to be consistent with existing inclusive mass bounds~\cite{Aaboud:2017leg,CMS:2018efh,Bagnaschi:2017tru} but provide experimentally feasible cross sections. To be concrete, we take four slepton mass points, 120, 200, 250 and 300 GeV, with in each case a relatively low mass splitting of $\Delta M = M_{\Sl} - M_{\widetilde{\chi}^0_1}=10$~GeV and 20~GeV, corresponding to the compressed scenario discussed above.

The di--lepton system of interest is defined by requiring an electron--positron
or muon--antimuon pair, with \pt\ $>$ 5~GeV and $|\eta| < 2.5$ (or
$|\eta| < 4.0)$ be present, ensuring that reasonable reconstruction and
trigger efficiencies may be expected. Here, the enlarged $\eta$-coverage case
corresponds to the ATLAS/CMS upgraded tracker for the High--Luminosity (HL)--LHC \cite{Collaboration:2017mtb,Klein:2017nke}. 
Here, and for all exclusive processes below, we use the \textsc{SuperChic} 2.07 Monte Carlo (MC)
generator~\cite{Harland-Lang:2015cta,SuperCHIC} (the results of the more recent version 3~\cite{Harland-Lang:2018iur} will be very similar since the version 3 is updated
to include heavy ion collisions but the aspects that deal with photon-initiated production in $pp$ collisions are essentially identical for the purposes of our study). We find that for such exclusive processes
the signal cross section is rather low, ranging from 0.03~fb to 0.8~fb for the
considered mass region. 

It is therefore essential to select these events during nominal LHC running, where the instantaneous luminosity and hence pile--up activity is relatively high. As we will see, the average number of pile--up events, $\mu$, will significantly
affect the size of the background, and with this in mind
we will consider three reference points, namely
$\mu = 0, 10$ and 50, with the $\mu = 0$ point only serving to
disentangle pile--up from non-pile--up effects. In each case we will consider a reference 
integrated luminosity of 300~\fbinv\,, which can be accumulated within a few years by ATLAS and CMS experiments
under current running conditions. After the HL--LHC upgrade collecting this amount of data
would occur in a matter of months, however here even higher values of $\mu$ will be present and the prospects for running with FPDs in this environment requires close
consideration, currently ongoing~\cite{HL-LHC}.
It should also be noted that at high $\mu$ expected at HL-LHC, keeping the rates
of L1 triggers based on low-\pt\ leptons low (and hence getting low pre-scales)
will be challenging. We assume that it will be possible to sufficiently reduce
them by constructing a trigger based on double-tagged protons and on topological
and other cuts which will be explained below.

At high $\mu$, the probability of no pile--up interactions is essentially zero and therefore we cannot select exclusive events by requiring rapidity gaps be present in the central detectors. Instead we will apply a $z$-vertex veto, which
requires that no vertices and tracks be present within $\pm$ 1~mm of the primary vertex.
This procedure has been shown  to be efficient in rejecting inclusive and pile--up
backgrounds, while maintaining reasonable signal selection efficiencies, in the ATLAS
 measurements of exclusive muon pair~\cite{Aaboud:2017oiq} and $WW$~\cite{Aaboud:2016dkv}
production, and the CMS + TOTEM measurement of exclusive muon pair production~\cite{Cms:2018het}.

The cuts we apply can be divided into
three classes, namely: cuts on the detected protons in the FPDs (`FPD 
  cuts'); requiring no other additional charged particles in the central detector
(`no-charged cuts'); and the selection applied to the lepton pair (`di--leptons cuts'). The FPD cuts are applied at generator level
by simply requiring both protons to be in a region where a sufficiently high
acceptance can be expected, namely $0.02 < \xi < 0.15$. The lepton cuts are
applied
at generator level too but detector inefficiencies are accounted for by applying
reconstruction efficiencies for electrons and muons as functions of their \pt.
We take overall efficiencies as found in the inclusive slepton search
\cite{Aaboud:2017leg}. This considers a similar lepton \pt\ range (in fact in \cite{Aaboud:2017leg} an even lower \pt\ $>4$ GeV cut is applied) and includes all physics effects, in contrast to specialized combined performance studies which usually split various effects and show
efficiencies for higher \pt. Nevertheless the lepton efficiencies taken from
these inclusive slepton searches may in fact be overly conservative
when applied to leptons in
exclusive processes, which are not generally accompanied by additional particles.
With this in mind we have applied the muon reconstruction efficiencies, which are
typically by 10--15\% higher than the electron ones, for both muons and
electrons in our study. These are to a good approximation flat in $\eta$ and
are not expected to vary strongly with the amount of pile--up.

\begin{table}
\begin{center}
\begin{tabular}{|c||c|c||} 
\hline
Efficiency & \multicolumn{2}{c||}{$\langle \mu \rangle_{PU}$} \\ 
\cline{2-3} 
 & 10  & 50 \\ \hline \hline
$z$-veto & 0.843 & 0.481 \\ \hline
vertex & 0.100 & 0.391 \\ \hline
tracks & 0.057 & 0.128 \\ \hline
\end{tabular}
\caption{\small{The signal $z$-vertex veto efficiency
  for the mass combination
  $M_{\tilde{l}}/M_{\widetilde{\chi}^0_{1}}$ = 200~GeV/180~GeV and two values of $\mu$. The `vertex' (`tracks')
  row shows the fractions of events where at least one vertex (where no vertex but
  at least one track) is found in the region $\pm$ 1~mm from the primary
  vertex.}}
\label{zveto}
\end{center}
\end{table}

On the other hand, the efficiency of the no-charged or $z$-vertex veto cuts
will vary with $\mu$, and so here we make use of the
Delphes framework \cite{deFavereau:2013fsa}, a software package providing a fast simulation
of detector response that takes
into account the effect of magnetic field, the granularity of the calorimeters
and sub-detector resolutions. Delphes is also able to overlay a specified number
of pile--up events with an existing MC signal or background event sample. 
For this study, input cards with ATLAS simulation parameters are supplied, while the pile--up events are generated with \textsc{Pythia} 8.2~\cite{Sjostrand:2014zea} in minimum bias (MB) mode.
Table~\ref{zveto} shows the $z$-vertex veto efficiency for one specific signal
sample defined by $M_{\tilde{l}}/M_{\widetilde{\chi}^0_{1}}$ = 200~GeV/180~GeV, but
similar values are observed for the other mass combinations. Here, and throughout this paper we will consider two pile--up scenarios, namely $\langle \mu \rangle=10,50$. While the latter is relevant for nominal LHC running, the former is taken to give an idea of the scaling of the various efficiencies and event numbers we consider with pile--up, although the precise value is not of direct experimental relevance. The efficiency is
defined in a sample of events containing at least two leptons with \pt\ $>$
5~GeV and $|\eta| < 2.5$, as a ratio of events that have no additional vertices and tracks
in the region of $\pm$ 1~mm around the primary vertex, to all selected
signal events. 
The $z$-vertex veto efficiencies are found to be in agreement with those
estimated in \cite{Aaboud:2017oiq,Aaboud:2016dkv}. For events that do not pass the
$z$-vertex veto requirements, the `vertex' row in table~\ref{zveto} shows the
fraction of those that had at least one vertex with $|z_{\rm vtx} - z_{\rm prim}| <$
1~mm and the `tracks' row shows the fraction of those
that had all vertices with $|z_{\rm vtx} - z_{\rm prim}| >$ 1~mm, but there was at least
one track with $|z_{\rm trk} - z_{\rm prim}| <$ 1~mm. Note that all ratios in one
column sum up to 100\%. 

All applied cuts in this analysis are summarized in
table~\ref{cuts}. Some of these are chosen in order to suppress specific background contributions, and
will be explained in the following sections. After applying all cuts specified
in table~\ref{cuts} and applying the lepton efficiencies as described above, the resulting signal event yields for an integrated luminosity
of 300~\fbinv\  are given in table~\ref{signal}. For completeness, we include here the single dissociation (SD) and double dissociation (DD) contributions, where one or two proton from the dissociation system registers in the FPDs; this will be discussed more later. Here we note that while the SD contribution reaches 6-9\%
for a mass of 120~GeV, they are below 1.5\% elsewhere and the DD contribution is
completely negligible. Enlarging the pseudorapidity range to $|\eta| < 4.0$
 increases  the signal yields by 10\% at most. 

\begin{table}
\begin{tabular}{||c|c|c||} 
\hline
&$5<p_{T,l_1,l_2} <$ 40~GeV &$|\eta_{l_1,l_2}| < 2.5$ (4.0) \\ \cline{2-3}
&Aco $\equiv 1-|\Delta\phi_{l_1l_2}|/\pi >$ 0.13 (0.095) &$2 < m_{l_ll_2} < 40$~GeV \\ \cline{2-3}
Di--lepton&$\Delta R (l_1,l_2) > 0.3$ & $|\eta_{l_1}-\eta_{l_2}| < 2.3$ \\ \cline{2-3}
&$\bar{\eta}\equiv |\eta_{l_1}+\eta_{l_2}|/2 < 1.0$&$||\vec{p_{Tl_1}}|-|\vec{p_{Tl_2}}|| > 1.5$~GeV\\ \cline{2-3}
& $W_{\rm miss} > 200$~GeV & \\ \hline\hline
FPD& $0.02 < \xi_{1,2} < 0.15$ &$p_{T,{\rm proton}} < 0.35$~GeV \\ \hline\hline
No--charge& No hadronic activity & z-veto \\ \hline
\end{tabular}
\caption{\small{Cuts used in this analysis.}}
\label{cuts}
\end{table}

\begin{table}
\begin{center}
\begin{tabular}{|c||c|c|c|c||} 
\hline
scenario & \multicolumn{4}{c||}{lepton \pt\ interval [GeV]} \\ 
\cline{2-5} 
$M_{\tilde{l}}/M_{\widetilde{\chi}^0_{1}}$ & 5---15 & 5---20 & 5---30 & 5---40 \\ \hline\hline
120/100 & 0.4 & 0.9 & 2.2 & 2.8 \\ \hline
120/110 & 1.2 & 2.4 & 3.7 & 3.9 \\ \hline
200/180 & 0.2 & 0.8 & 1.9 & 2.2 \\ \hline
200/190 & 1.4 & 1.9 & 2.3 & 2.3 \\ \hline
250/230 & 0.1 & 0.4 & 1.1 & 1.2 \\ \hline
250/240 & 0.8 & 1.1 & 1.2 & 1.2 \\ \hline
300/280 & 0.1 & 0.2 & 0.6 & 0.7 \\ \hline
300/290 & 0.4 & 0.6 & 0.6 & 0.6 \\ \hline
\end{tabular}
\caption{\small{Signal event yields for integrated luminosity of 300~\fbinv\
    for four
  mass assignments and four lepton \pt\ intervals after applying all cuts
  specified in table~\ref{cuts}. Yields correspond to the sum of $\mu^+\mu^-$ and $e^+e^-$ final states
  observed in $|\eta| < 2.5$ and to the sum of left and right-handed sleptons.
  Lepton reconstruction efficiencies taken from~\cite{Aaboud:2017leg} as well as
  SD and DD contributions are included.}}
\label{signal}
\end{center}
\end{table}

\section{Photon--initiated backgrounds}\label{QEDbg}
\subsection{$\gamma\gamma\to W^+W^-\to l^+\nu +l^-\bar{\nu}$}

\begin{figure}
\includegraphics[height=.29\textheight,width=0.49\textwidth]{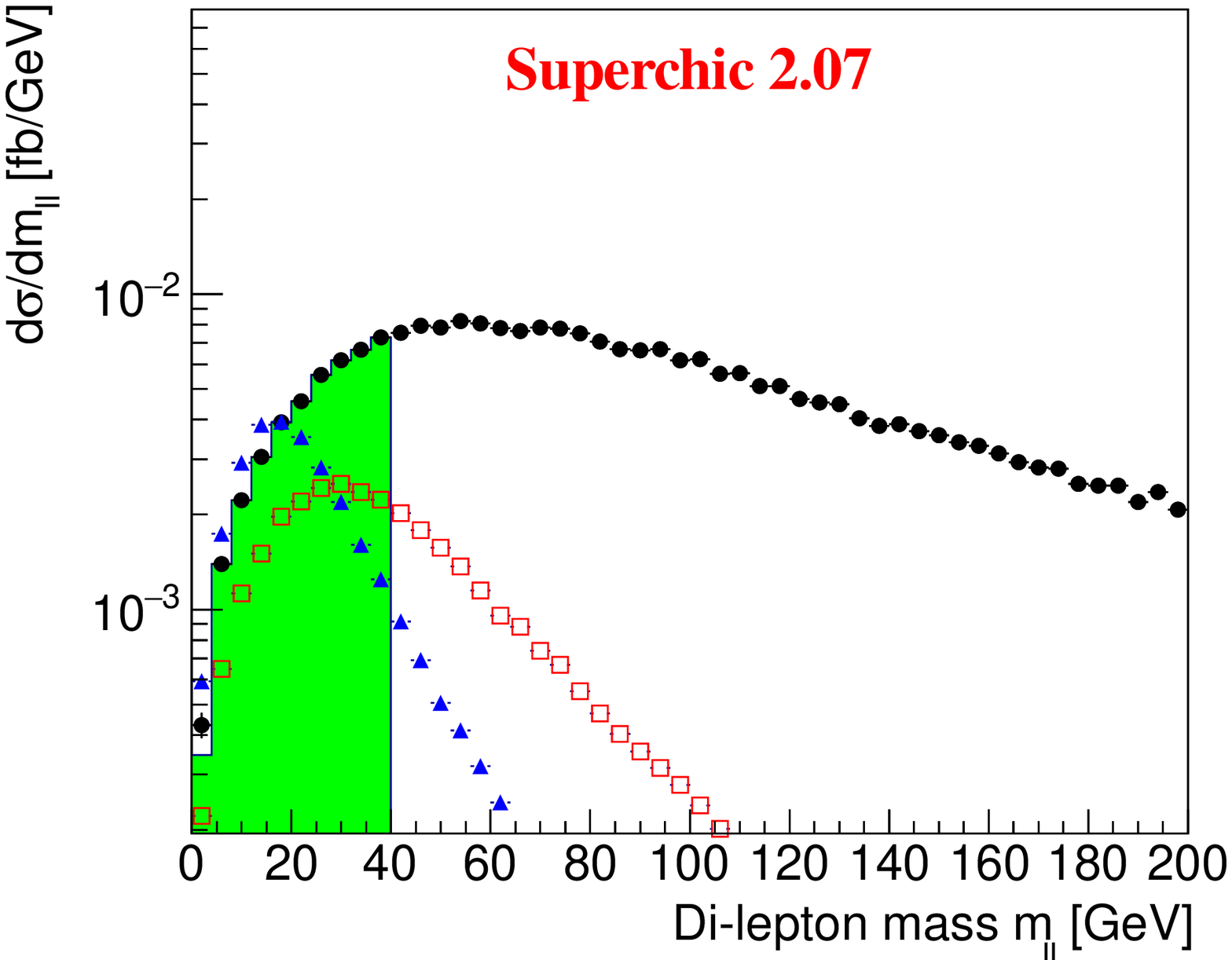}
\includegraphics[height=.29\textheight,width=0.49\textwidth]{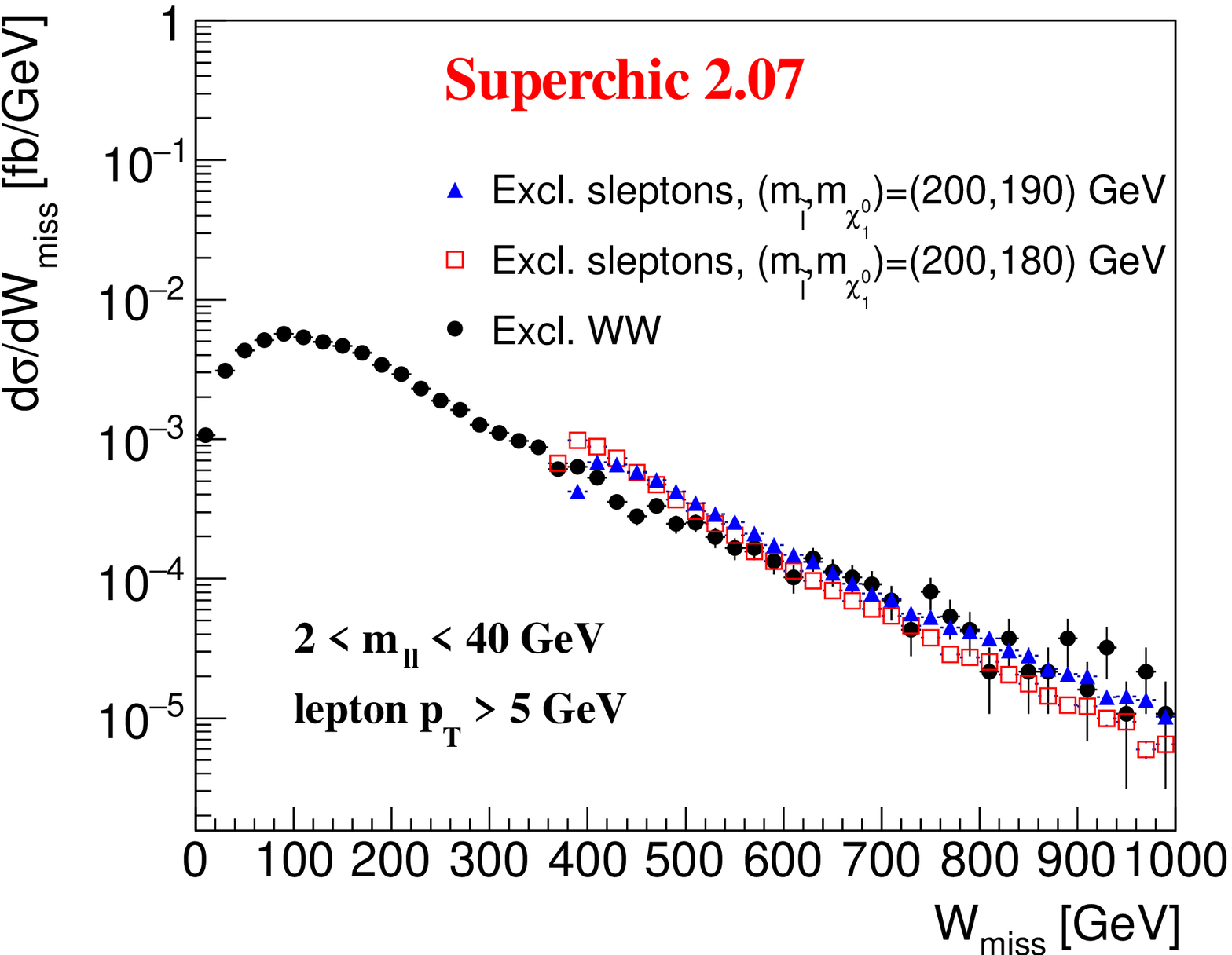}  
\caption{\small{Distribution of di--lepton mass (left) and missing mass $W_{\rm miss}$ (right) for
    exclusive slepton pair and $WW$  production. The di--lepton distribution
  was obtained before applying any cuts and the green area indicates the
  acceptance region used in the analysis. The $W_{\rm miss}$ distribution was
  obtained after applying lepton \pt\ and di--lepton mass cuts, see
  table~\ref{cuts}.}}
\label{ss-WW}
\end{figure}

The production of a $WW$ pair followed by leptonic decays via the same photon--initiated production mechanism as the signal is one of the major sources of background.
Here, the production cross section via the combined $e^+e^-$ and $\mu^+\mu^-$ decay channels is about 1~fb prior to any final--state cuts, and so is somewhat larger than the signal.
However, here the lepton \pt\ is peaked at $\sim M_W/2$, in contrast to the signal, which prefers lower values. We therefore place a $2 < m_{l_ll_2} < 40$~GeV cut on the di--lepton invariant mass and a related kinematic cut of $p_{T,l_1,l_2} < 40$ GeV, significantly reducing this background. An additional cut on the missing mass $W_{\rm miss} >$ 200~GeV, constructed from the momenta of the protons in the FPDs and of the leptons in the central detector, reduces this background further. This in particular corresponds to the invariant mass of the neutrino (neutralino) pair in the exclusive $WW$ (SUSY) cases, and therefore for the signal we must have $W_{\rm miss} > 2 m_{\widetilde{\chi}^0_{1}}$, while the background is peaked at much lower values.
Distributions of these two variables for the signal and the exclusive
$WW$ background are compared in figure~\ref{ss-WW}. The di--lepton mass is plotted
for all events before applying any cuts and the $W_{\rm miss}$ distribution is shown
after applying cuts on the lepton \pt\ and di--lepton mass. As expected, both demonstrate a good
potential in suppressing this background.

The expected numbers of background events after applying the cuts from table~\ref{cuts} for
leptons in the $|\eta| < 2.5$ region are summarized in table~\ref{WW}. Very
similar numbers are obtained for $|\eta| < 4.0$. Because of the
large central system masses, which must be larger than the $W_{\rm miss}$ cut of 200 GeV, the SD and DD dissociation contributions (see the following section for definitions) are negligible, as the corresponding
$\xi$ values fall out of the FPD acceptance. For our final results, we will consider the first
row. The other rows would become relevant in the case of an observation, in which case by 
 tuning the $W_{\rm miss}$ cut threshold such that $W_{\rm miss} > 2 M_{\widetilde{\chi}^0_{1}}$, we
would be able to suppress this background more strongly.

\begin{table}
\begin{center}
\begin{tabular}{|c||c|c|c|c||} 
\hline
$W_{\rm miss} >$ X & \multicolumn{4}{c||}{lepton \pt\ interval [GeV]} \\ 
\cline{2-5} 
[GeV] & 5---15 & 5---20 & 5---30 & 5---40 \\ \hline\hline
X = 200 & 0.04 & 0.2 & 0.5 & 0.6 \\ \hline
X = 400 & 0.03 & 0.1 & 0.3 & 0.3 \\ \hline
X = 500 & 0.03 & 0.07 & 0.2 & 0.2 \\ \hline
X = 600 & 0.02 & 0.05 & 0.1 & 0.1 \\ \hline
\end{tabular}
\caption{\small{Event yields at 300~\fbinv\ for 
  exclusive $WW$ production via leptonic decays, for four $W_{\rm miss}$ cut values
  (corresponding to the four mass scenarios considered in this study) and four
  lepton \pt\ intervals, after applying all cuts specified in
  table~\ref{cuts}. Yields
  correspond to the sum of $\mu^+\mu^-$ and $e^+e^-$ final states observed in
  $|\eta| < 2.5$. Lepton reconstruction efficiencies taken from
  \cite{Aaboud:2017leg}.}}
    \label{WW}
    \end{center}
\end{table}

Finally, we note that the above situation is much worse for the case of
chargino pair production. Though an obvious advantage of this process is the
possibility to search in the final state for a pair of different flavour
leptons (e.g. $e^-$ and $\mu^+$), which allows to exclude the QED background
from low mass lepton production (discussed below), here, the branching ratio for
the leptonic decay is approximately 5 times smaller. After accounting for the
larger (up to factor of 4) production cross section in the chargino case (see
e.g.~\cite{HarlandLang:2011ih}), we find that the expected signal will be about
7---9 times smaller, and so the $WW$ background will strongly exceed the
signal.

\subsection{Low mass $\gamma\gamma\to l^+l^-$ production}

The exclusive production of lepton pairs in the FPD acceptance region, i.e. with $m_{l_ll_2}>$ 280 GeV, has a very small production cross section and will moreover lead to extremely different final--state lepton kinematics in comparison to the signal. This will therefore not provide a source of background. On the other hand, the production cross section for lepton pairs in the $m_{l_ll_2}>$10 GeV region, corresponding to the \pt\ $>$ 5~GeV cuts in the central detector, is about 8.4~pb, that is many orders of magnitude higher than the signal. While purely exclusive production will in this case not produce protons in the FPD acceptance, the situation changes if we allow the outgoing protons to dissociate, that is consider semi--exclusive production. Here, a proton from the dissociation system may still be detected in the FPD, but with a $\xi$ value that no longer matches the rapidity and mass of the central system. While the probability that this occurs is very small, nonetheless in combination with the significantly larger di--lepton production cross section, the resulting background may be rather large. To calculate this probability, we can either apply an analytic expression for the leading proton distribution, as fit to soft hadronic data, or directly evaluate this using appropriate event samples generated with \textsc{Pythia} 8.2. As discussed further in appendix~\ref{treg}, these give very similar results. We in particular calculate this probability to be $P_{\rm SDnel} \approx 0.7\%$ (where the subscript indicates that this is due to protons from the inelastic side of
a SD event) for a proton to lie in the FPD acceptance, which while small is in fact significantly larger than the suppression in the signal cross section relative to the $m_{l_ll_2}>$ 10 GeV di--lepton cross section. We therefore have to apply further cuts to suppress this source of background.

The cross sections for semi--exclusive lepton pair production were calculated by applying the procedure outlined in~\cite{Harland-Lang:2016apc}. That is, we supplement the result of  \textsc{SuperChic} 2.07 with  `effective' photon fluxes given according to elastic and low/high scale inelastic photon emission from the proton, but which pass the additional veto on central particle production due to the no-charged requirement. Both SD and DD contributions, where only one or both protons dissociate, respectively, are included. Further details are given in appendix~\ref{flux}. We consider contributions from electron, muon and $\tau$ pair production, with the latter case followed by leptonic decays simulated by \textsc{Pythia} 8.2.

How can we reduce this background?
First, we note that in the SD case, in order for a relatively low mass di--lepton to be produced with a sufficiently large $\xi$ value on the elastic proton side, the lepton pair must be produced at forward rapidity in the proton direction.
Therefore, to suppress this contribution we require $\bar\eta=|\eta_{l_1}+\eta_{l_2}|/2 < 1$.  Furthermore, a $|\eta_{l_1}-\eta_{l_2}|<2.3$ cut rejects events with a
large value of $m_{l_ll_2}$ but a rather small $p_{Tl}$. However, after
applying these cuts the
low-mass $\gamma\gamma\to l^+l^-$ background still exceeds the expected signal.
To further reduce the background, due both to SD and DD, we introduce a cut on transverse momentum of the forward proton, $p_{T, {\rm proton}}<0.35$~GeV, as the $p_T$ of the proton produced in the dissociation system will generally be larger
than that in the elastic case. This cut is applied directly in our calculation of appendix~\ref{treg}, and reduces the probability for a proton from the dissociation system to be registered in the FPD significantly, to $P_{\rm SDnel} \approx 0.4\%$.
In addition, in exclusive production the transverse momenta carried by the initial--state photons are generally very small and hence to good approximation the leptons are produced back--to--back, with equal and opposite transverse momenta, that is with close to zero acoplanarity.
While allowing for proton dissociation will generally increase the average photon $q_T$, and hence wash this out somewhat, nonetheless the background is significantly more peaked at low acoplanarity in comparison to the signal, where the leptons from the slepton decay can be produced at arbitrary $\phi$. We therefore apply an additional acoplanarity cut, $ 1- |\Delta\phi_{l_1l_2}|/\pi  > 0.13 \,(0.095)$, for leptons in the $|\eta| < 2.5 \, (|\eta| < 4.0)$ range. We in addition apply a cut on the difference $|p_{Tl_1}-p_{Tl_2}| > 1.5$~GeV (where $p_{Tl_i} = |\vec{p_{Tl_i}}|$), which reduces the background for the same reason. The method for applying these cuts is explained in more detail in appendix~\ref{flux}. 

The result of imposing these cuts is summarized in table~\ref{mumu}. 
For the SD case, we find that the lower $\xi$ threshold shifts the di--lepton
mass spectrum to higher values and in combination with the $\bar{\eta}$ cut the
onset of the di--lepton mass spectrum moves beyond the 40~GeV threshold for electron/muon production. For $\tau$-pair production the di--lepton masses are on average lower
but nevertheless we find that the SD background
is in all cases under control, while in the DD case it is found to be negligible, see
table~\ref{mumu}. This is due to the impact of the branching
ratio for the decay to a lighter lepton, and because
the two neutrinos from a $\tau$ decay carry away a significant fraction of the $\tau$ energy, such that only
18\% of leptons from the decay survive the lepton \pt\ cut. 
On the other hand, we find a non--negligible contamination  from electron and muon pair production via
DD interactions. 

\begin{table}\begin{center}
\begin{tabular}{|c|c||c|c|c|c||} 
\hline
Exclusive & Proton & \multicolumn{4}{c||}{lepton \pt\ interval [GeV]} \\ 
\cline{3-6} 
di--lepton & dissociation & 5---15 & 5---20 & 5---30 & 5---40 \\ \hline\hline
$e^+e^- + \mu^+\mu^-$ & SD & $\sim$0/$\sim$0 & $\sim$0/$\sim$0 & $\sim$0/$\sim$0 & $\sim$0/$\sim$0 \\
\cline{2-6}
& DD & 1.4/1.1 & 1.4/1.1 & 1.4/1.1 & 1.4/1.1 \\ \hline
$\tau^+\tau^-$ & SD & 0.01/0.00 & 0.03/0.02 & 0.05/0.02 & 0.05/0.02 \\ 
\cline{2-6}
& DD & $\sim$0/$\sim$0 & $\sim$0/$\sim$0 & $\sim$0/$\sim$0 & $\sim$0/$\sim$0 \\ \hline
\end{tabular}
\caption{\small{Event yields for an integrated luminosity of 300~\fbinv\ for 
  di--lepton production in four lepton \pt\ intervals after applying all cuts
  specified in table~\ref{cuts}. Results for single and double proton dissociation are given, and with
 $|\eta|<2.5$ / $|\eta|<4.0$ intervals for the final--state lepton. Lepton
  reconstruction efficiencies are taken from~\cite{Aaboud:2017leg}. The purely exclusive contribution is exactly zero due to the mass acceptance of the FPDs and cuts imposed on $m_{l_ll_2}$, so are not shown. The values marked as $\sim$0
correspond to numbers which are sufficiently below 0.01.}}
    \label{mumu}
    \end{center}
\end{table}

\section{QCD induced exclusive pair production}\label{CEPbg}

QCD--induced exclusive hadron production with leptonic decays may also mimic the signal final--state. We consider three classes of this background below, namely vector meson production, $K^+K^-$ pair production and $D^+ D^-$ pair production.

In the case of vector meson ($\rho$, $J/\psi$ etc) production with leptonic decays, these backgrounds can be straightforwardly removed by omitting the regions of lepton pair invariant mass corresponding to the meson masses.

The CEP of $K^+K^-$ pairs requires a more careful treatment, but we again find this to be negligible. In particular, the \textsc{SuperChic} 2.07 prediction for this process, with $m_{KK} > 10$ GeV and kaon \pt\ $>$ 4~GeV is 1.3 fb, that is of the order of the signal. However, as in the case of photon--initiated lepton pair production, protons from the purely exclusive process will not register in the FPDs, and so we must again consider SD and DD semi--exclusive production. Once we include the small probabilities for protons from the dissociation system to be registered, we find that this background is very small. This will be further reduced by the fact that not all kaons will decay in the central detector due to the non--negligible kaon path length.

A potentially more significant source of background is due to the production of
heavy mesons which have a reasonable branching to leptons plus neutrals, e.g.
$D^+\to\mu^+\nu\pi^0$. While the CEP of $D^+D^-$ pairs is not currently generated
in \textsc{SuperChic} 2.07, the CEP of $c\bar{c}$ pairs, a natural source of
$D$-mesons, is implemented, and hence we use this to estimate the size of the
corresponding background. The cross section for masses of the $c\bar{c}$ system
$m_{c\bar{c}}>$~10 GeV and quark \pt\ $>$ 5~GeV is found to be about 3~nb. In addition, we consider the production of a low mass $gg$ pair, which while having a very low probability to produce mesons which decay in such a way as to mimic the signal, nonetheless has a very large production cross section. We in particular find that this amounts to $\sim 2$ $\mu$b for the same mass and \pt\ cuts applied to the parton--level cross section as in the $c\bar{c}$ case. We note that as the elastic and SD cross sections are
measured to be roughly the same at HERA~\cite{Khoze:2006gg}, we do not modify
the gluon fluxes to account for dissociation and have simply used the cross
sections given by \textsc{SuperChic} 2.07 for the elastic case.
Again, for these low masses it is only SD and DD events followed by proton hits in the FPDs from the dissociation system that can pass the event selection, and this requirement will reduce the backgrounds considerably.
Indeed, we find that after applying cuts from table~\ref{cuts}, the expected number of
events for luminosity of 300 \fbinv\ is below 0.01 in all cases, and hence is
under very good control.

\section{Non-diffractive events with pile--up}\label{ND}

Leptons with relatively low \pt\ are produced copiously at LHC: the inclusive
production cross section is about 10~nb for \pt\ $>$ 5~GeV~\cite{Bolzoni:2012kx}. The main source of these is from semi-leptonic decays of heavy
mesons (e.g. $D^+\to\mu^+\nu\pi^0$) and the decay of $W$-bosons, while the decay
of kaons and pions inside the central detector  will also contribute~\cite{ATLAS:2011uea}. If such an inclusive event coincides with protons registered in the
FPDs due to unrelated pile--up interactions, this can contribute as a
considerable source of background. To evaluate the effect of this we generate
the dominant source of inclusive lepton production, due to non--diffractive (ND)
jet production, with both \textsc{Pythia} 8.2~\cite{Sjostrand:2014zea} and \textsc{Herwig} 7.1~\cite{Bahr:2008pv,Bellm:2015jjp}. We find that the cross sections for jet \pt\ $>$ 7~GeV (taken as a
conservative lower value) are unsurprisingly very large, with \textsc{Herwig} 7.1
(\textsc{Pythia} 8.2) giving 16 (27)~mb. Note that here we have fully included
initial--state radiation (ISR), final--state radiation (FSR) and  multi-parton
interactions (MPI) in order to completely model the background, including the
underlying event (UE).

From a practical point of view, this cross section is so much larger than the signal ($\sim 0.1$ fb) that it is impossible to generate a sufficiently large initial background sample in order to accurately evaluate the impact of cuts that reduce the background to the signal level, that is by roughly 14 orders of magnitude. To bypass this issue, we have instead considered the three approximately factorized classes of cuts introduced in section~\ref{sig}, see table~\ref{cuts}. In particular, we consider these cuts to be sufficiently independent that the final rejection efficiencies can be taken as the product of those due to applying the cuts individually. We discuss the impact of these in turn below.

\subsection{FPD cuts}

We first evaluate the probability $P_{FPD}$, defined by the ratio of `double tag'
(DT) events containing a proton passing the FPD cuts in table~\ref{cuts} on each
side of the interaction point and passing the ToF criteria discussed below, to
the total event number. As the probability of registering a proton in the FPDs
increases with $\mu$, this will clearly be quite sensitive to the total amount
of pile--up. Note that these protons are dominantly due to soft SD events, with
the most significant source of background then being due
to an overlay of three pile--up events, i.e. two soft SD events which each
produce a proton on one side of the interaction point and a hard-scale ND event
that produces two leptons which pass all other cuts. However, there is some
contribution from soft ND events, which may also produce protons in the FPD
acceptance, and therefore we use MB events (including both diffractive and non--diffractive contributions) to evaluate
the total effect of pile--up interactions. 

At high pile-up the probability of such MB events giving hits in the FPDs is significant, while the probability of the final--state in the central detector from the
hard-scale ND event mimicking the signal is small but non--zero. It is therefore
challenging to suppress this background to a tolerable level. However, an
effective way to do this is to equip the FPDs with time-of-flight (ToF)
detectors. Then, by measuring precisely the arrival times of the two individual
protons on each side, one is able to suppress the pile--up background
significantly. In general for the signal the $z$-vertex coordinate calculated from the proton arrival time in the FPDs will coincide with the vertex position of the central event within the vertex resolution, whereas for the
background these will not. The designed time resolution is
10~ps~\cite{AFP,CT-PPS} for both AFP and CT-PPS, and much progress has been made
to achieve this already. 

\begin{figure}
\begin{center}
\includegraphics[height=.28\textheight,width=0.49\textwidth]{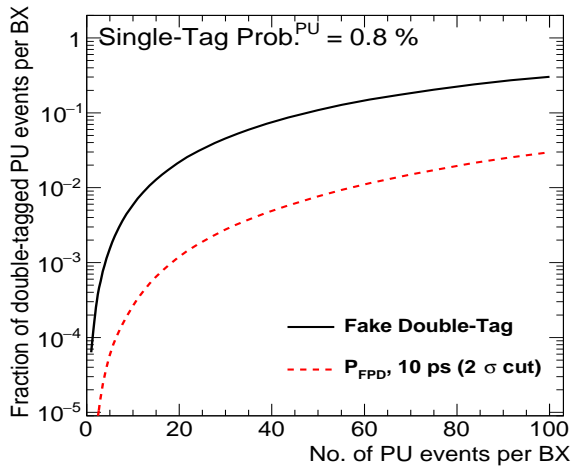}
\caption{\small{Fractions of fake DT events in one bunch crossing (BX) and overall rejection probabilities
    $P_{FPD}$, including ToF discrimination, as a function of $\mu$. The
    results are based on a ST probability of 0.008 for MB events generated by
    \textsc{Pythia} 8.2 and on a ToF resolution of 10~ps.}}
\label{fakeDT}
\end{center}
\end{figure}

In the results below we will therefore include this rejection factor, calculated
by requiring that the $z$ position of the central vertex coincides with that
calculated from the ToF detectors within this $\sigma_t=$ 10~ps resolution. In
more detail, we assume that the bunch longitudinal width is 7.5~cm for
non--zero $\mu$, and we require that the difference between the two arrival
times be within $2\sigma_t$. Under these assumptions, we then generate ten
million bunch crossings for each $\mu$ point, with a number of MB events in
 each case given by a Poisson distribution and with the
$z$-coordinate of the primary vertex given by a Gaussian distribution. 
Considering all DT events within this sample, and calculating the corresponding
difference in arrival times of the proton hits, we then calculate the fraction
of events where the evaluated vertex lies within $\sigma_t$ (corresponding to
3~mm) on either side of the primary vertex. These results are summarized in
figure~\ref{fakeDT}. Note that the ToF reduction factors linearly depend on the
inverse ToF resolution (for more details, see \cite{Tasevsky:2014cpa,Cox:2007sw}).

The calculated probabilities for a fake DT event to occur, and the ToF rejection
factors are shown in table~\ref{rejFPD}. The total probability $P_{FPD}$ is
then given by the fake DT probability divided by the ToF rejection factor. 
We show results for both \textsc{Pythia} 8.2 and \textsc{Herwig} 7.1, and with two different values of
$\mu$, and can see that the total rejection varies from 2 to 4 orders of
magnitude, depending on the pile--up and generator used. 
\begin{table}[h]
\begin{center}
\begin{tabular}{|c||c|c||c|c||} 
\hline
& \multicolumn{2}{c||}{\textsc{Pythia} 8.2} & \multicolumn{2}{c||}{\textsc{Herwig} 7.1}\\ 
\cline{2-5} 
& \multicolumn{2}{c||}{$\langle \mu \rangle_{PU}$} & \multicolumn{2}{c||}{$\langle \mu \rangle_{PU}$}\\  
\cline{2-5} 
& 10 & 50 & 10 & 50 \\ \hline
Fake DT & 0.0048 & 0.105 & 0.0123 & 0.222 \\
ToF rejection & 18.3 & 13.7 & 17.5 & 11.3 \\
$P_{FPD}$ & 2.6 $\times 10^{-4}$ & 7.6 $\times 10^{-3}$ & 7.0 $\times 10^{-4}$ & 2.0 $\times 10^{-2}$ \\ \hline
\end{tabular}
\caption{\small{Probabilities of fake double-tagged events, rejection factors
    due to ToF detectors and overall probabilities to observe in one bunch
    crossing two forward protons in the FPDs and an inelastic vertex in the
    central detector consistent with that obtained by ToF detectors within
    the time resolution of 10~ps.
  }
  }
\label{rejFPD}
\end{center}
\end{table}

We can see that the results vary by a factor of $\sim 2$ depending on whether
\textsc{Pythia} 8.2 or \textsc{Herwig} 7.1 is used. Considering the probability that a proton from an
individual MB ('single-tag', ST) event passes the FPD cuts in
table~\ref{cuts}, we find a value of 0.008 (0.013) for \textsc{Pythia} 8.2 (\textsc{Herwig} 7.1).
If we add the combinatoric effect of pile--up, for example for $\mu = 50$, and
use Poisson statistics, this probability increases to the value
$(1-(1-0.008)^{49}) \approx$ 0.34. The square of that, i.e. the fake DT
probability, would then be 0.116, in a good agreement with the exact value of
0.105 obtained by the procedure based on generating large samples of MB events
described above.
The overall probability $P_{FPD}$ can then be approximately obtained by
considering that the vertex must lie within a 6~mm window, in comparison to the
total available region of 15~cm, and then accounting for the combination of
different arrival times between the two FPDs that may be consistent with the
primary vertex gives, for $\mu = 50$ and \textsc{Pythia} 8.2, an additional factor of 1.5.
In total, the expected overall rejection factor is then $0.11\times 0.6\times 1.5 /15 \sim 0.0066$, roughly consistent with the more precise results in
table~\ref{rejFPD}.

More generally, the simulation of the inelastic pile--up events that lead to
the DT hit are fundamentally due to soft QCD, and therefore there will clearly
be some model dependence in these results. As a first check, we extract the ST
probability, but with MPI turned off, and find that the impact is rather small
increasing it by $\sim 10\%$.
On the other hand the $\xi$ spectra of the leading protons at LHC energies have
not been tuned in the MC, as there is currently no LHC data available to do
this. As a consistency check we therefore extract the ST (at $\mu=0$)
probability at \sq\ = 1.8~TeV, and find a value of 0.023 (0.024) for \textsc{Pythia} 8.2
(\textsc{Herwig} 7.1), which are both consistent with CDF data~\cite{Goulianos:1998wt} and
with the prediction based on the triple-Regge vertex~\cite{Kaidalov:1973tc}
developed to describe the ISR data, as well as the ABMST model prediction~\cite{Appleby:2016ask}. Thus at these lower energies where tuning has been performed
for the $\xi$ spectra the two MCs agree with each and with theoretical
expectations. We therefore conclude that the ST($\mu=0$) probability is driven
mainly by the overall energy scale, rather than the strength of MPI.
In the future  we may expect an improvement in the accuracy of the predictions
at the LHC energies, as the models are refined and tuned to further LHC data. 

\subsection{No charged cuts}
The probability to see no charged particles in the central detector other than
two leptons for non-zero pile--up values is evaluated using
the following relation:
\begin{equation}
P_{\rm no-ch}(\mu=10,50) = P_{\rm gap}(\mu=0)\times P_{z-{\rm veto}}(\mu=10,50)
\label{Pgap}
\end{equation} 
where $P_{\rm gap}$ is estimated at particle level, while for $P_{z-{\rm veto}}$ the
values from table~\ref{zveto} are used. The latter was calculated using signal
events generated by \textsc{SuperChic} 2.07, hadronized by \textsc{Pythia} 8.2 and with the ATLAS detector
response simulated by Delphes. We note that the probability $P_{\rm gap}(\mu=0)$
is not exactly zero in non--diffractive events (see for example~\cite{Khoze:2010by}) as gaps may be
generated by fluctuations in hadronization. Nonetheless, it  is expected to be
rather small. As the number of events with two leptons coming from background
processes is very small, see below, we estimate $P_{gap}(\mu=0)$ using events
selected so as to resemble the signal as closely as possible. In particular, we
consider event samples with at least two charged particles with \pt\ $>$ 5~GeV
and $|\eta| < 2.5 (4.0)$, where two of them are isolated by $\Delta R > 0.3$,
separately for each of the background processes with sufficiently high cross
sections, namely $c\bar{c}$ and $gg$ CEP and inclusive ND jet production.
We then evaluate $P_{\rm gap}(\mu=0)$ as a fraction of events in this sample
which have no additional charged particles with \pt\ $>$ 0.4~GeV and
$|\eta| < 2.5 (4.0)$. Track reconstruction efficiencies in these $p_T$ and
$\eta$ regions are satisfactory. The values of $P_{\rm no-ch}$ for these processes and three
values of pile--up are shown in table~\ref{rejnoch}. 

In principle, $P_{\rm no-ch}$  can be obtained directly from the Delphes
simulation by estimating the $P_{z-{\rm veto}}$ factors. For the three background processes,
we got the same fraction of events rejected by finding at least one pile--up
vertex within 1~mm of the primary vertex as for the signal,
namely 10\% at $\mu = 10$ and 39\% at $\mu = 50$ (see the `vertex' row of
table~\ref{zveto}) and finally we obtained even lower values of $P_{z-{\rm veto}}$ than
$P_{\rm no-ch}$. However, to be conservative we quote the higher
values, as the primary vertex reconstruction algorithm implemented in Delphes
requires further testing, in particular for the very high values of $\mu$ envisaged
for the HL-LHC where it is not expected to distinguish all vertices and rather to
merge some of them, especially in the region of beam spot (where the density
of vertices is highest, see e.g. \cite{Aaboud:2016rmg}).

\begin{table}
\begin{center}
\begin{tabular}{|c||c|c|c||} 
\hline
$P_{\rm no-ch}$ & \multicolumn{3}{c||}{$\langle \mu \rangle_{PU}$} \\ 
\cline{2-4} 
    &     0            &        10     & 50         \\ \hline\hline
CEP $c\bar{c}$ & $3.5\times10^{-3}$ & $2.9\times10^{-3}$ & $1.7\times10^{-3}$ \\ \hline
CEP $gg$       & $3.3\times10^{-5}$ & $2.8\times10^{-5}$ & $1.6\times10^{-5}$ \\ \hline
Incl. jets $(|\eta|<2.5)$    & $5.2 (2.0)\times10^{-7}$ & $4.4  (1.7)\times10^{-7}$ & $2.5 (1.0)\times10^{-7}$     \\ \hline
Incl. jets $(|\eta|<4.0)$    & $1.7 (0.7)\times10^{-7}$ & $1.4 (0.6)\times10^{-7}$ & $0.8 (0.3)\times10^{-7}$     \\ \hline
\end{tabular}
\caption{\small{The no-charged rejection probabilities as a function of $\mu$ for  $c\bar{c}$ and $gg$ CEP, and inclusive
  ND jet production.
   The numbers in the first column were obtained at particle level
  and then used to calculate the numbers in the other columns using eq.~\ref{Pgap}
  and $P_{z-veto}$ probabilities from table~\ref{zveto}. The inclusive jet events were
 generated with \textsc{Pythia} 8.2 (\textsc{Herwig} 7.1}).}
 \label{rejnoch}
  \end{center}
\end{table}

In this context it is worth noting that if the $\eta$-coverage of the
central tracker would be enlarged as approved for the ATLAS (so called ITk
\cite{Collaboration:2017mtb}) and CMS \cite{Klein:2017nke} upgrades for Run
III, the probability $P_{\rm gap}$ would decrease. Indeed, applying the same
procedure as above and only requiring no charged particles in the range
\pt\ $>$ 0.4~GeV and $|\eta| < 4.0$, the probability $P_{gap}$ for inclusive events drops by a factor
of three - as can be seen by comparing the inclusive jet rows in
table~\ref{rejnoch}. 

\subsection{Di--lepton cuts}

 To evaluate the impact of the di--lepton cuts we select a sample of events where exactly two same--flavour and opposite sign leptons
 (electrons or muons) are produced with \pt\ $>$ 5~GeV and $|\eta| < 2.5$ and
 satisfying most of the di--lepton cuts of table~\ref{cuts}.
  Furthermore, we reject those events
  where the two selected leptons are accompanied by charged particles from the
  same heavy-meson decays. In other words, an event is rejected if at least one
  charged particle (besides the di--lepton pair) coming from the same decay as the
  lepton from the di--lepton pair has \pt\ $>$ 0.4~GeV and $|\eta| < 2.5$. This
  effectively rejects decays modes with extra charged particles such as e.g.
  $D^0\to K^-e^+\nu$ or $D^+\to\rho^0\mu^+\nu$. 
  
  The probability $P_{lep}$ is then
  defined as the fraction of events surviving these cuts, and including the lepton reconstruction
  efficiencies from \cite{Aaboud:2017leg}.
 We find $P_{lep}=0.8\times10^{-7}$ ($2.5\times10^{-7}$)  with \textsc{Pythia} 8.2 (\textsc{Herwig} 7.1).\footnote{Note that,
 as  \textsc{Pythia} 8.2  does not include W bosons in inclusive ND jets, we have estimated this 
  contribution with \textsc{Herwig} 7.1, see section \ref{FY}.}
As mentioned in section~\ref{sig}, these reconstruction
efficiencies are not expected to depend significantly on the amount of pile--up,
so the $P_{lep}$ rejection factors were obtained at generator level only and with $\mu=0$, but can safely be used up to $\mu = 50$.

\subsection{Comparisons at zero pile--up}

It is useful to compare and discuss the above results at zero pile--up, in order
to further elucidate the differences between the two generators. The total jet
cross sections, $P_{FPD}$, $P_{gap}$ and $P_{lep}$ are all estimated as explained
above using a sample of ND inclusive jet events with \pt\ $>$ 7~GeV. The ratios
of the \textsc{Herwig}~7.1 to \textsc{Pythia}~8.2 results  are found to be
0.67, 0.55, 0.40 and 3.3, respectively. We can therefore see that
\textsc{Herwig}~7.1 has a higher particle multiplicity in inclusive ND jet
events in comparison to \textsc{Pythia}~8.2. In more detail we find that it
produces roughly 4 times as many events with at least two charged hadrons with
\pt\ $>$ 5~GeV and $|\eta| < 2.5$ and about twice as many events with a
di--lepton pair with \pt\ $>$ 5~GeV and $|\eta| < 2.5$. For this reason, we
find that $P_{gap}$ is almost by a factor of 2.5 times lower in \textsc{Herwig}~7.1 in comparison to \textsc{Pythia} 8.2.

Since \textsc{Pythia}~8.2 does not include the $W$ boson production in ND jets
the value of $P_{lep}$ in PYTHIA is by a factor of 3.3 times lower.
The $P_{lep}$ in PYTHIA is then corrected using the fraction of $W$ boson events
in all di--lepton events found in HERWIG.

For our purposes, and in particular given the very small sizes of all background
probabilities $P_{FPD}$, $P_{gap}$ and $P_{lep}$, in some cases based on
quantities which were not yet tuned at LHC (e.g. large rapidity gaps or $\xi$
spectrum of the leading proton), these differences are acceptable.

\subsection{Inclusive non-diffractive background at zero pile--up}
A final source of background we have to in general consider is due to events where the proton hits in the FPDs are produced by a single underlying inelastic interaction, which also produces a lepton pair in the central detector. Such a background would equally be present at $\mu=0$, as well as for realistic $\mu$ values. The  probability, $P_{acc}$, that this occurs is given by the product
 \begin{equation}
 \label{mu0}
 P_{acc}=P'_{FPD}\cdot P_{lep}\cdot P_{gap}\ ,
\end{equation}  
 where $P'_{FPD}$ was calculated using \textsc{Pythia} 8.2 including MPI, from the
 inclusive jet sample generated with \pt\ $>$ 7~GeV, where at least two charged
 particles are observed with \pt\ $>$ 5~GeV and $|\eta| < 2.5$. We then
 calculate the fraction of events in which a proton in the FPD acceptance of
 table~\ref{cuts} on one side is observed. The total probability is given by
 squaring this. For a mixture of ND and SD events with a dynamically generated
 values of the soft survival probablity, $S^2$ (a quantity which is available
 from \textsc{Pythia} 8.2), we find $P'_{FPD}=(0.0004)^2$. 
 Since the cross section for dijet production is $\sigma(p_T>7\mbox{GeV})=27$~mb
 and $P_{lep}=0.8\times 10^{-7}$ and $P_{gap}=5.2\times 10^{-7}$, the expected
 background is negligible. 

\section{Results}\label{FY}

\begin{table}[h]
\begin{center}
\begin{tabular}{|l||c|c|c||} 
  \hline
Event yields / & \multicolumn{3}{c||}{$\langle \mu \rangle_{PU}$} \\ 
\cline{2-4} 
$\cal L = $ 300~\fbinv\ & 0 & 10  & 50 \\ \hline\hline
Excl. sleptons & 0.6---3.9 & 0.5---3.3 & 0.3---1.9 \\ \hline
Excl. $l^+l^-$  & 1.4 & 1.2 & 0.7 \\ \hline
Excl. $K^+K^-$  & $\sim 0$ & $\sim 0$ & $\sim 0$ \\ \hline
Excl. $W^+W^-$  & 0.7 & 0.6 & 0.3 \\ \hline
Excl. $c\bar{c}$& $\sim 0$ & $\sim 0$ & $\sim 0$ \\ \hline
Excl. $gg$      & $\sim 0$ & $\sim 0$ & $\sim 0$ \\ \hline
Incl. ND jets & $\sim 0$($\sim 0$) & 0.1(0.1) & 1.8(2.4)  \\ \hline
\end{tabular}
\caption{\small{Final event yields corresponding to an integrated luminosity of
  300~\fbinv\ as a function of amount of pile--up events per bunch
  crossing for the slepton signal and all considered background processes. 
  All numbers correspond to the di--lepton mass range
  $2 < m_{l_1l_2} < 40$~GeV and lepton 5 $<$ \pt\ $<$ 40~GeV and a tracker coverage of
  $|\eta| <2.5$. The ranges
  in the signal event yields illustrate the
  spread obtained from the entire studied slepton mass range: the lower value
  comes from the $(M_{\Sl},M_{\widetilde{\chi}^0_{1}}) = (300,280)$~GeV, the higher from the $(M_{\Sl},M_{\widetilde{\chi}^0_{1}}) = (120, 110)$~GeV
  scenario. The value marked as $\sim$0 corresponds to a number which is
  sufficiently below 0.01. The inclusive ND jet events were
 generated with \textsc{Pythia} 8.2 (\textsc{Herwig} 7.1).}}
\label{AllEY2.5}
\end{center}
\end{table}

We collect our results for the expected signal and background event yields in tables~\ref{AllEY2.5} and~\ref{AllEY4}. Here, the former case corresponds to $|\eta| < 2.5$ (i.e. the current tracker coverage) while the latter corresponds to $|\eta| < 4.0$ (i.e.
the upgraded tracker coverage). To give a global picture, these results correspond to the full di--lepton mass range of $2 < m_{l_1l_2} < 40$~GeV, although
information about individual lepton \pt\ ranges for processes where it is
relevant can be found in tables~\ref{signal},~\ref{WW} and~\ref{mumu}. 
In summary, we observe that in total 2--3 signal events for 300~\fbinv  \,can be
expected, with a $S/B \sim 1$. We note that \textsc{Pythia} 8.2 and
(\textsc{Herwig} 7.1) give similar predictions for the contamination from the
inclusive ND jets. 
 These relatively small numbers therefore clearly do not correspond to a statistically significant observation.
There are however various ways to improve this situation.


\begin{table}
\begin{center}
\begin{tabular}{|l||c|c|c||} 
  \hline
Event yields / & \multicolumn{3}{c||}{$\langle \mu \rangle_{PU}$} \\ 
\cline{2-4} 
$\cal L = $ 300~\fbinv\ & 0 & 10  & 50 \\ \hline\hline
Excl. sleptons & 0.7---4.3 & 0.6---3.6 & 0.3---2.1 \\ \hline
Excl. $l^+l^-$  & 1.1 & 0.9 & 0.5 \\ \hline
Excl. $K^+K^-$  & $\sim 0$ & $\sim 0$ & $\sim 0$ \\ \hline
Excl. $W^+W^-$  & 0.6 & 0.5 & 0.3 \\ \hline
Excl. $c\bar{c}$& $\sim 0$ & $\sim 0$ & $\sim 0$ \\ \hline
Excl. $gg$      & $\sim 0$ & $\sim 0$ & $\sim 0$ \\ \hline
Incl. ND jets   & $\sim 0$($\sim 0$) & 0.03(0.05) & 0.6(0.7)  \\ \hline
\end{tabular}
\caption{\small{The same as in table~\ref{AllEY2.5} but for the enlarged tracker
  coverage $|\eta|<4.0$.}}
\label{AllEY4}
\end{center}
\end{table}

From the point of view of the phenomenological analysis presented here, the situation may be improved by cutting on the variable proposed in~\cite{HarlandLang:2011ih}, namely the maximum kinematically allowed values of $m_{\tilde{\chi}}$ and $m_{\tilde{l}}$ assuming the signal decay chain.
Following the approach of ~\cite{HarlandLang:2011ih}, we have checked that these cuts lead to some mild improvement in the signal significance over the background, though the interplay between these and other cuts on $W_{\rm miss}$ and $m_{l_1l_2}$, $p_{T,l}$ is rather delicate and requires some fine--tuning. We may also expect some reduction in the low mass dilepton SD and DD
backgrounds, but in the absence of a full MC implementation this cannot
currently be calculated.

 Experimentally, the signal yield can be
 doubled by taking all di--lepton masses into account. This would, however,
 not only increase the background but also the average di--lepton
 mass itself and hence limit the possibility of estimating the
 unknown mass of the DM particle by measuring the central system mass via the  FPDs. 
 Another way to increase the signal yield would be to increase the lepton
 reconstruction efficiencies. In the ATLAS study \cite{Aaboud:2017leg} they start
 at 70\% for muons and at 50\% for electrons and slowly rise with increasing
 \pt, nevertheless since we deal with two leptons, any increase in the 
 single-lepton efficiency could have a reasonable impact.

 The background contamination, in turn, could be lowered by considering the
 following points. Since a big part of background leptons comes from decays of
 heavier particles, they originate from a vertex displaced from the primary one.
 It would thus be natural to consider restricting track longitudinal, $z_0$,
 and transverse, $d_0$, impact parameters to some small
 values for all found leptons in background samples (as is done for example when
 determining reconstruction efficiencies of primary vertices, see
 \cite{Aaboud:2016rmg}). Actual LHC $b$--physics analyses rather use a pseudo-proper
 lifetime as a more appropriate observable to separate primary from secondary
 vertices. Furthermore, as discussed above, both ATLAS and CMS are
 upgrading their trackers to cover the additional region
 $2.4 < |\eta| < 4$. Both are also considering adding timing detectors in these
 forward areas with resolution of about 30~ps \cite{CMSMIP,ATLASHGTD}
 The time information for tracks
 in this forward area will improve the track-to-vertex association, leading to
 a performance similar to that in the central region for both jet and lepton
 reconstruction, as well as the tagging of heavy-flavour jets.
 The timing in the central detector will allow us to check that not only the
longitudinal, $z$, coordinate but also the {\em time} of lepton pair emission,
measured via the forward protons ToF detectors, 
coincides (or not in the PU case) with the values measured in the central
detector. That is we acquire another ToF rejection factor in addition to that
shown in table~\ref{rejFPD}.

Finally, the background from events with
proton dissociation may in principle be suppressed by rejecting events with
particles observed in the proton dissociation region, i.e. at large rapidity, with a good timing resolution potentially allowing these particles to be distinguished from those that originate from pile--up. Unfortunately, as it stands the detector coverage of ATLAS or CMS is not sufficient in this region to achieve this but building Zero Degree Calorimeters designed to be radiation hard and providing also timing information~\cite{Murray} is under discussion both in ATLAS and CMS.

Before concluding, it is worth comparing in more detail with the results of the alternative study presented in~\cite{Beresford:2018pbt}. We note that here only those backgrounds due to genuinely exclusive production, dominantly $W^+W^-$ pair production, are considered, while pile--up effects, ND, SD and DD production are not accounted for. Some care is therefore needed in making such a comparison. Considering the experimentally most promising scenario, $(M_{\Sl},M_{\widetilde{\chi}^0_{1}}) = (120, 110)$~GeV, and taking for the sake of direct comparison the no pile--up scenario, we can see from Table~\ref{AllEY2.5} that we expect 3.9 signal events and 0.7 $W^+ W^-$ background events, corresponding to a $\sim 1.5 (2.6) \,\sigma$ significance for $100 (300)$ ${\rm fb}^{-1}$. From Fig. 3 of~\cite{Beresford:2018pbt} we can see that a somewhat larger $\sim 2\,\sigma$ significance at $100$ ${\rm fb}^{-1}$ is found. This is however entirely expected: our numbers presented even for the experimentally unrealistic no pile--up scenario include various cuts, as in Table~\ref{cuts}, that are imposed to reduce the SD and DD backgrounds and which will correspondingly reduce this signal significance. Indeed, removing these increases the signal significance such that it is indeed rather consistent with~\cite{Beresford:2018pbt}. However, clearly this is not the experimentally relevant case, and we should instead take the high pile--up case in  Table~\ref{AllEY2.5}, and include the corresponding formally reducible SD, DD and ND jet backgrounds. In this case the significance is clearly lower, but nonetheless this presents a more realistic picture of the current situation.

\section{Conclusions}\label{sum}

In this paper we have discussed the prospects for searching for slepton pair production via leptonic decays in compressed mass scenarios at the LHC, via photon--initiated production. In this case the experimental signal is simple, comprising only four charged particles in the final state, namely two forward outgoing protons and two leptons in the central detector. 
 Theoretically, such models are well motivated by cosmological and naturalness considerations as well as $(g-2)_{\mu}$ phenomenology. Moreover, in inclusive channels the production of electroweakinos in such compressed mass scenarios is highly challenging due to the low production cross sections and high backgrounds. The exclusive search channel therefore offers a potentially unique method to probe such challenging regions of SUSY parameter space. Furthermore, it has the advantage that the predicted production cross sections are model independent, being driven only by the electric charge of the produced states.

On the other hand, the predicted production cross sections are small, being $\sim 0.1$  fb or less after accounting for all relevant cuts and efficiencies, depending on the
slepton mass in the experimentally allowed region. Therefore it is essential to collect these events at nominal LHC luminosities and for any backgrounds to be under very good control. 
In this work we have analysed in detail all major sources of background under these conditions, where in particular pile--up will be high. In more detail, we have considered: the irreducible photon--initiated $WW$ background; the reducible background from the semi--exclusive photon--initiated production of lepton pairs and QCD--initiated production of gluon and $c$--quark jets (via leptonic decays of hadrons produced in hadronization) at low mass, where a proton produced in the initial proton dissociation registers in the forward proton detectors; the reducible pile--up background where (dominantly) two independent single--diffractive events coincide with an inelastic lepton pair production event. For the proton dissociation and pile--up backgrounds we have performed dedicated MC simulations, including most of relevant detector effects and efficiencies, in order to evaluate their impact as accurately as possible.

We have found that requiring that the lepton pair lie in the signal $m_{l_1l_2}<40 $ GeV region, combined with further judicially chosen cuts on the lepton momenta
 leads to significant reductions in the background. The pile--up backgrounds are strongly reduced by the use of fast timing in the proton tagging detectors, which is certainly essential  to perform such a measurement, as well as the aforementioned lepton cuts and a further cut on the proton transverse momentum. These also help to reduce the semi--exclusive backgrounds considerably. However, after accounting for all of these effects we find that the backgrounds from pile--up and semi--exclusive photon--initiated lepton pair production are nonetheless expected to be of the same order as the signal, with the irreducible $WW$ background being somewhat lower. 
 
 Nonetheless, the intention of our study is to be as comprehensive and to a certain extent as conservative as possible. We have therefore also discussed a variety of ways in which this situation could be improved upon, with the potential for increased tracker acceptance combined with timing detectors at the HL--LHC being particularly promising. While a detailed study of the possibilities at the HL--LHC is beyond the scope of the current work, this provides a strong motivation for further work on this area, and for collecting data with tagged protons there. Certainly the main backgrounds are in principle reducible and therefore with further investigation we may be able to reduce these further.
The discussed search strategy could also be used to explore other simplified
models for Dark Matter with small mass splitting between the DM and its charged
co-annihilation partner.

\section{Acknowledgements}
We thank Torbj\"{o}rn Sj\"{o}strand, Kasuki Sakurai, David Cerde\~{n}o, Mike
Albrow, Alexander
Kup\v{c}o, Pavel \v{R}ezn\'{i}\v{c}ek, Mike Hance, Zbyn\v{e}k Dr\'{a}sal,
Michal Mar\v{c}i\v{s}ovsk\'{y} and Michael Murray for
useful discussions. This work was supported by the Ministry of
Education, Youth and Sports of the Czech Republic under the project LG15052.
MGR thanks the IPPP at the University of Durham for hospitality.
VAK acknowledges the support from the Royal Society of Edinburgh Auber award.

\appendix
\section{Coherent, incoherent and evolution contributions to photon flux}\label{flux}
 
 In this appendix we discuss in more detail how the case of semi--exclusive photon--initiated production is accounted for. In particular the contribution to the photon flux from elastic emission ($p\to p+\gamma$), incoherent emission  ($p\to N^*/\Delta+\gamma$) and DGLAP emission from quarks should be considered. While the incoherent is dominant for lower mass proton dissociation, the DGLAP contribution dominates at higher mass. All these contributions can be precisely accounted according to the procedure outlined in~\cite{Manohar:2016nzj}, however this corresponds to the purely inclusive case, whereas here we require that no particles be produced within the central detector acceptance. To achieve this, we apply the approach of~\cite{Harland-Lang:2016apc}, to calculate `effective' photon fluxes which survive an additional veto on central particle production.
 
We are then interested in three classes of events, namely when both photons are emitted
coherently (`elastic'), one is emitted coherently and
the other incoherently (`single dissociation', SD) and  both are
emitted incoherently (`double dissociation', DD). In the default version of \textsc{SuperChic} 2.07 only the first, purely elastic case is generated.
Thus while this case is always treated exactly, such that cuts on the final--state particles can be directly imposed, in the latter two cases an effective approach must be taken. In particular, photon fluxes due to incoherent and DGLAP emission are calculated as in~\cite{Harland-Lang:2016apc} with a veto imposed on additional particle production in the relevant $\eta$ region, and with the acoplanarity cut imposed by limiting the $z,Q^2$ integral (see (13) of~\cite{Harland-Lang:2016apc})
\begin{equation}
q_T = \left(Q^2(1-z)\right)^{1/2}> \pi\cdot {\rm Aco}\cdot p_T\;,
\end{equation}
which generates the photon flux.  For simplicity, we take $ p_T \sim \frac{m_{l_1l_2}}{2}$ for the lepton transverse momentum, but note that a more precise treatment could give a somewhat different value for the final flux.

For the case of the cut on the scalar
difference $|p_{Tl_1}-p_{Tl_2}|$, the photon $q_T^2$ distribution
was generated logarithmically  between $q_0 = 0.5$ GeV and $m_{l_1 l_2} /2$, where the lower scale is chosen so as to include low mass incoherent emission. This is then translated into the lepton momenta by including the appropriate angular integration.

In both cases, these are then multiplied by average survival factors of 0.85 and 0.15 in the SD and DD cases, respectively.

 \section{Probability for proton hit via dissociation}\label{treg}

 To calculate the probability that a proton produced in the original proton dissociation system registers in the FPDs, we have evaluated the fraction of SD events generated by \textsc{Pythia} 8.2 where a proton on the dissociation side is found within $0.02 < \xi < 0.15$ region:
\begin{equation}
P_{\rm SD,nel} \simeq 0.007\;.
\end{equation}
If in addition, the proton is required to have $p_T < 0.35$~GeV, it reduces to
\begin{equation}
P_{\rm SD,nel} \simeq 0.004\;.
\end{equation}
Alternatively, if we consider the fraction of ND events, with MPI turned off, which from Feynman scaling arguments we may also expect to give some estimate of this probability, we find
\begin{equation}
P_{\rm ND} \simeq 0.005\;.
\end{equation}
We note that one can also apply the tools of Regge theory to get an analytic
estimate of $P_{\rm SD,nel}$. In particular, we use the
triple-Regge formula for the soft inclusive proton-proton cross section 
\begin{equation}
\label{3R}
\frac{d \sigma}{d^2p_T d \xi}=\sum_{i,j}G_{iij}(p^2_T)(1-\xi)^{1-2\alpha_i(p^2_T)}\;,
\end{equation}
where $G_{iij}$ is the triple Reggeon vertex, and the sum is over all contributing Reggeons, with the requirement that $i\neq I\!P$, i.e. the Pomeron exchange contribution, which corresponds to the elastic case where the proton does not dissociate, is excluded.

We in particular apply the fit of~\cite{Kaidalov:1973tc}, assuming that the leading proton distribution at the
LHC can be described by the same triple-Regge parameterization as this fit, which was tuned to
describe CERN-ISR data (note that the more recent fit of~\cite{Appleby:2016ask} gives rather similar results). The probability to observe the proton in the FPD is then simply given by
\begin{equation}
\label{P3R}
P_{\rm SD,nel}=\frac 1{\sigma_{inel}}\int d\xi d^2p_T\frac{d\sigma}{d^2p_T d\xi}\ . 
\end{equation}
where the integral is performed over the relevant $\xi$ and $p_T$ region. For $0.02 < \xi <0.15$ we find
\begin{equation}
P_{\rm SD,nel}\simeq 0.01\;,
\end{equation}
while imposing \pt\ $< 0.35$ GeV in addition gives
\begin{equation}\label{eq:psd}
P_{\rm SD,nel} \simeq 0.0046 \;.
\end{equation}

Thus this value of $P_{\rm SD,nel}$ is in remarkable agreement with the MC value
$P_{\rm SD,nel}$, and the latter is used throughout.
 
  \bibliography{arxiv}{}
\bibliographystyle{JHEP}
  
\end{document}